\documentclass[a4paper,11pt]{article}
\usepackage{jheppub}
\usepackage[T1]{fontenc} 
\usepackage{amsmath}
\usepackage[caption=false]{subfig}

\usepackage{breqn}

\newcommand{\dd}{{\rm d}}
\newcommand{\nn}{\nonumber}

\newcommand{\ep}{\epsilon}
\newcommand{\matr}{\mathrm}

\allowdisplaybreaks[4]

\title{Evaluating the last missing ingredient for the three-loop quark static potential by differential equations}

\author[a]{Roman N. Lee,}
\author[b]{Vladimir A.\ Smirnov}
\affiliation[a]{Budker Institute of Nuclear Physics,
630090 Novosibirsk, Russia}
\affiliation[b]{Skobeltsyn Institute of Nuclear Physics of Moscow State
University, 119992 Moscow, Russia}
\emailAdd {r.n.lee@inp.nsk.su}
\emailAdd {smirnov@theory.sinp.msu.ru}

\abstract{
We analytically evaluate the three-loop Feynman integral which was the last missing ingredient for the analytical evaluation of the three-loop quark static potential.
To evaluate the integral we introduce an auxiliary parameter $y$, which corresponds to the residual energy in some of the HQET propagators.
We construct a differential system for 109 master integrals depending on $y$ and fix boundary conditions from the asymptotic behaviour in the limit $y\to \infty$. The original integral is recovered from the limit $y\to 0$.
To solve these linear differential equations we try to find an $\ep$-form of the differential system. Though this step appears to be, strictly speaking,
not possible, we succeed to find an $\ep$-form of all irreducible diagonal blocks, which is sufficient for solving the differential system in terms of an $\ep$ expansion.
We find a solution up to weight six in terms of multiple polylogarithms and obtain an analytical result for the required three-loop Feynman integral by taking the limit $y\to 0$. As a by-product, we obtain analytical results for some Feynman integrals typical for
HQET.}


\begin{document}

\maketitle
\flushbottom

\section{Introduction}

Three-loop corrections to the static potential between two heavy quarks\footnote{See Ref.~ \cite{tlpana} for a brief overview on calculations of corrections to the static quark potential and their applications.} were evaluated in Refs.~\cite{Smirnov:2008pn,Smirnov:2009fh,Anzai:2009tm}.
However, these results involved three constants which were obtained only in a numerical form.
In order to find these constants in an analytical form one might try to apply the DRA approach formulated in Ref. \cite{Lee:2009dh}.
An important step in this direction was made in Ref. \cite{Lee:2012te}. And indeed, the analytical results for the two of the three missing constants were obtained withing this approach and presented in Ref.~\cite{tlpana}. However the calculation of the last constant appeared to be much more involved for the DRA method, mostly, due to overwhelming numerical issues.

The goal of the present paper is to attack the problem from a different direction, by using the differential equations approach, along the lines of
Ref.~\cite{Henn:2013nsa}. We successfully apply this strategy thus finding the last missing ingredient for a
completely analytical result for the three-loop static quark
potential. The result for the potential is presented in Ref.~\cite{tlpana}.

The three-loop Feynman integral which up to now was considered as the most complicated
ingredient for the analytical evaluation of the three-loop quark static potential is
$I_1=F_{1,\ldots,1,-2,1,0}$, where
\begin{multline}
F_{a_1,\ldots,a_{12}}=\int\int\int\frac{\dd^D k\;\dd^D l\;\dd^D r}
{(-k^2)^{a_1}(-l^2)^{a_2}(-r^2)^{a_3}(-(r+q)^2)^{a_4}(-(k-l+r+q)^2)^{a_{5}}(-(k+q)^2)^{a_6}}
\\ \hspace*{-0mm} \times
\frac{(-v\cdot l)^{-a_{10}}(-(k-r)^2)^{-a_{12}}}
{(-(l-r)^2)^{a_7}(-(k-l)^2)^{a_8}(-v\cdot k)^{a_9}  (-v\cdot r)^{a_{11}}}\;,
\label{fam12ind}
\end{multline}
with the usual $-i0$ implied in all propagators, $v\cdot q=0$, is a non-planar family of dimensionally regularized (with $D=4-2\ep$) Feynman integrals with twelve indices.
As the dependence on both $v^2$ and $q^2$ is trivially factorized, we put $v^2=-q^2=1$ in what follows.
There is only one master integral in the sector with $a_i>0, i=1,2,\ldots,9,11$, $a_{10},a_{12}\leq 0$ and it is more
convenient \cite{Smirnov:2010zc} to choose it to be $I_1$,
rather than the corner integral $F_{1,\ldots,1,0,1,0}$ of this sector.
The integral $I_1=I_1(\ep)$ is finite in four dimensions, $D=4$.
The numerical value  $I_1(0)\approx-20.9484$ was obtained in the calculation of Refs.~\cite{Smirnov:2008pn,Smirnov:2009fh}
(see also \cite{Smirnov:2010zc}) using a Mellin-Barnes representation.

In order to use the method of differential equations
\cite{Kotikov:1990kg,Kotikov:1991pm,Remiddi:1997ny,Gehrmann:1999as,Gehrmann:2000zt,Gehrmann:2001ck,Henn:2013pwa,Lee:2014ioa} one has to introduce one auxiliary
parameter \cite{Henn:2013nsa}. Though it might look like unnecessary complication, the idea behind this step is the following.
An auxiliary parameter, when chosen judiciously, allows one to "deform" the original family of integrals to the point which is
more accessible for other methods, in particular, to DRA method. Then, putting boundary conditions at this point and using powerful
machinery of the method of differential equations, one can obtain results for the original integrals.

In the next section we explain how we introduce an auxiliary parameter
and solve differential equations for the corresponding set of the master integrals.
In Section~3 we describe how boundary conditions are fixed. As a by-product,
we obtain analytical results for a family of Feynman integrals typical for
the Heavy Quark Effective theory (HQET).

\section{Master integrals and differential equations}

In order to introduce an auxiliary parameter, one might think of relaxing the condition $v\cdot q = 0$. However, this choice complicates the situation considerably because one effectively obtains three scales, or two variables in differential equations.
Instead, we introduce parameter $y$ by the following replacement
\begin{equation}
\frac{1}{(-v\cdot k)^{a_9}} \to \frac{1}{(y/2-v\cdot k)^{a_9}}\,,
\end{equation}
and similarly for propagator number 11.
This prescription corresponds to the introduction of the residual energy in HQET propagators.
The factor $1/2$ is introduced for convenience.
We arrive at the following family of Feynman integrals
\begin{multline}
{F}_{a_1,\ldots,a_{12}}=\iiint\frac{\dd^D k\;\dd^D l\;\dd^D r}
{(-k^2)^{a_1}(-l^2)^{a_2}(-r^2)^{a_3}(-(r+q)^2)^{a_4}(-(k-l+r+q)^2)^{a_{5}}(-(k+q)^2)^{a_6}}
\\ \times
\frac{(-v\cdot l)^{-a_{10}} (-(k-r)^2)^{-a_{12}}}
{(-(l-r)^2)^{a_7}(-(k-l)^2)^{a_8}(y/2-v\cdot k)^{a_9} (y/2-v\cdot r)^{a_{11}}}\;.
\label{fam12ind-y}
\end{multline}

Using {\tt FIRE}  \cite{Smirnov:2008iw,Smirnov:2013dia,Smirnov:2014hma} combined with {\tt LiteRed} \cite{Lee:2012cn,Lee:2013mka}
we reveal 109 master integrals.
The derivation of differential equations for a family of master integrals is a straightforward procedure.
We take derivatives of the master integrals in $y$ with the help of {\tt LiteRed}  \cite{Lee:2012cn,Lee:2013mka} and then apply {\tt FIRE} to reduce the resulting integrals to master integrals. As a result we obtain a system of linear differential equations
\begin{eqnarray}
\partial_y \mathbf{F} &=& \matr A(y,\ep) \mathbf{F}\;,
\label{primaryde}
\end{eqnarray}
where $\mathbf{F}$ is the column-vector of primary master integrals and $\matr A$ is a $109\times 109$-matrix.

According to the strategy suggested in \cite{Henn:2013pwa} and first applied in \cite{Henn:2013tua,Henn:2013woa,Henn:2013nsa} and then in many other papers,
in particular at the four-loop level \cite{Henn:2016men},
it is reasonable to try to pass to a new basis (canonical basis in terms of Ref. \cite{Henn:2013pwa})
where the differential equations take the form with a factorized $\ep$-dependence on the right-hand side (or $\ep$-form for brevity).

To do this we use the algorithm  \cite{Lee:2014ioa} introduced by one of the author of the present paper. We successfully obtain a globally Fuchsian
form and an $\epsilon$-form for each diagonal block (which basically correspond to sectors). However, we fail to perform
the last step (namely, the factorization of the $\epsilon$-dependence) for the matrix as a whole. Therefore, we arrive at the form
\begin{eqnarray}
\partial_y \widetilde{\mathbf{F}} &=& \left(\ep \widetilde{\matr A}(y)+\widetilde{\matr B}(y,\ep)\right)\widetilde{\mathbf{F}}\;,
\label{almostcanonicalde}
\end{eqnarray}
where $\widetilde{\matr A}$ is a block-diagonal matrix independent of $\ep$ and $\widetilde{\matr B}$ is the strictly lower-triangular
matrix with zeros in each diagonal block of $\tilde{A}$. Besides, they both have a global Fuchsian form
\begin{equation}
\tilde{\matr A}(y)=\sum_k \frac{a_k}{y-y_k}\,,\quad
\tilde{\matr B}(y,\ep)=\sum_k \frac{b_k(\ep)}{y-y_k}
\label{Atildematrices}
\end{equation}
with $y_k$ running over the set $\{-1,-1/2,0,1,1/2\}$. Since we want to obtain the boundary conditions from the limit $y\to\infty$, we reduce the matrix residue at infinity, $-\sum_k (\ep a_k+b_k(\ep))$, to a Jordan normal form, which, in particular, means that we secure that
\begin{equation}\label{eq:zerosum}
\sum_k b_k(\ep)=0\,.
\end{equation}
Although the resulting system \eqref{almostcanonicalde} is not in an
$\ep$-form, this form is completely sufficient for our purposes. Indeed, for integrals of a certain sector, we have the system
\begin{equation}
\partial_y \widetilde{\mathbf{F}}_1 = \ep \tilde{\matr A}_1(y) \widetilde{\mathbf{F}}_1+\mathbf{R}(y,\ep)\;,
\end{equation}
where the inhomogeneous term $\mathbf{R}(y,\ep)$ is a linear combination of simpler master integrals, the coefficients being rational functions of $y$ and $\ep$. We assume that these simpler integrals have been already calculated at this point in a sense that we know their expansion in $\ep$ at fixed $y$ up to a sufficiently high order as well as their asymptotic behavior at fixed $\ep$ and $y\to y_k$. Then we also know the expansion and the asymptotic behavior of $\mathbf{R}(y,\ep)$, irrespectively on whether the coefficients of the linear combination are proportional to $\ep$. We only note that, the condition \eqref{eq:zerosum} essentially simplifies finding the asymptotics of $\mathbf{R}(y,\ep)$ at $y\to\infty$.

To solve the linear system (\ref{almostcanonicalde})  in a power expansion in $\ep$ we pass to the variable $x=1/y$ and obtain results in terms of multiple polylogarithms of the argument $x$, up to expansion coefficients of $109$ unknown functions $C_i(\ep)$.
The multiple polylogarithms \cite{Goncharov:1998kja} are defined recursively by
\begin{equation}\label{eq:Mult_PolyLog_def}
 G(a_1,\ldots,a_n;z)=\,\int_0^z\,\frac{\dd t}{t-a_1}\,G(a_2,\ldots,a_n;t)
\end{equation}
with  $a_i, z\in \mathbb{C}$ and $G(;z)=1$. In the special case where  $a_i=0$ for all i
one has by definition
\begin{equation}
G(0,\ldots,0;x) = \frac{1}{n!}\,\ln^n x \;.
\end{equation}
The letters $a_i$ for our result belong to the alphabet $\{-2,-1,0,1,2\}$.

\section{Fixing boundary conditions}

To fix the unknown constants in the result, we consider the leading order asymptotic behaviour of the solution of Eqs.~\eqref{almostcanonicalde} in the limit $y \to \infty$.
Terms of the corresponding expansion can be described in the language of expansion by regions \cite{Beneke:1997zp,Smirnov:2002pj}.
In our case, all the contributions of regions are obtained by considering each loop momentum to be
hard (i.e. $\sim y$) or soft (i.e. $\sim y^0$).
The crucial point in the matching is that the leading-order contributions are classified according to the power dependence on $y$.
This parameter enters with powers of the form $-k\ep$ where $k\in\{0,2,4,6\}$
and this is seen both from the point of view of differential equations and expansion by regions.

To fix the constants in our solutions of differential equations we
evaluate, first, the leading asymptotic behaviour associated with the region where all three loop momenta are hard, i.e. $\sim y$,
All these contributions are described as integrals of the following family
\begin{multline}
H_{a_1,\ldots,a_{9}}=\iiint\frac{\dd^D k\;\dd^D l\;\dd^D r}
{(-k^2)^{a_1}(-l^2)^{a_2}(-r^2)^{a_3}(-(k-l+r)^2)^{a_{4}}}
\\ \hspace*{-0mm} \times
\frac{(-v\cdot l)^{-a_{8}}}
{(-(l-r)^2)^{a_5}(-(k-l)^2)^{a_6}(y/2-v\cdot k)^{a_7}
  (y/2-v\cdot r)^{a_{9}}}\;.
\label{fam_hhh}
\end{multline}
We imply that $a_8\leq 0$. These integrals are typical to the HQET.

Using {\tt FIRE} we reveal 9 master integrals in this family,
\begin{eqnarray}
J_1=H_{0, 0, 1, 0, 1, 1, 1, 0, 0}\,,\;
J_2=H_{0, 0, 0, 1, 1, 1, 1, 0, 1}\,,\;
J_3=H_{0, 0, 1, 1, 1, 0, 1, 0, 1}\,,\;
J_4=H_{0, 1, 0, 1, 0, 1, 1, 0, 1}\,,\;
\nn \\  && \hspace*{-146mm}
J_5=H_{0, 1, 0, 1, 1, 1, 0, 0, 1} \,,\;
J_6=  H_{0, 1, 1, 1, 0, 0, 1, 0, 1} \,,\;
J_7=  H_{0, 1, 0, 1, 1, 1, 1, 0, 1} \,,\;
\nn \\  && \hspace*{-146mm}
J_8=  H_{1, 0, 1, 0, 1, 1, 1, 0, 1} \,,\;
J_9=  H_{1, 1, 1, 1, 0, 0, 1, 0, 1}\,.
\label{hhh_MI}
\end{eqnarray}

The first 5 master integrals can easily be evaluated by a consecutive integration over loop momenta and
expressed in terms of gamma functions at general $\ep$.
To evaluate the four more complicated integrals we apply the DRA method \cite{Lee:2009dh}
and use the dedicated package {\tt SummerTime} \cite{Lee:2015eva} for the calculation of multiple sums appearing underway.

In addition to the information about this region, we use
the leading order behaviour for
the region where all three loop momenta are soft, i.e. $\sim y^0$. The corresponding
contributions are either zero or expressed in terms of well-known massless propagator integrals.
It turns out that after this it is sufficient to add information about `intermediate' contributions
for a small number of master integrals to fix all the constants.

Upon fixing all the constants in our solution we obtain analytic results for all the 109 elements of the canonical basis from which we derive analytical results for the primary basis.
We checked our results numerically  with {\tt FIESTA} \cite{Smirnov:2015mct}.

The integral $I_1=F_{1,\ldots,1,-2,1,0}$ at $\ep=0$  can be obtained as the value at $y=0$ of
the corresponding integral $\bar{I_1}(y)$
depending on $y$ and belonging to the family (\ref{fam12ind-y}). Since we obtained our results for the 109 master integrals in terms of multiple polylogarithms with the variable $x=1/y$ we obtain also an expression for  $\bar{I_1}(1/x)$ in a similar form.
We used two ways to analytically evaluate the limiting values at  $x\to \infty$.
First, we evaluate $\bar{I_1}(1/x)$ at the point $x=10^{200}$ with
the computer implementation \cite{GiNaC_code} of {\tt GiNaC} \cite{GiNaC}
with the precision of 200 digits and this happens to be enough
to arrive at an analytical result using the {\tt PSLQ} algorithm \cite{PSLQ}.
Second, we rewrite the result for  $\bar{I_1}(y)$ in terms of multiple polylogarithms of the variable $y$ using our implementation of an algorithm presented in Ref.~\cite{GiNaC},
then set $y=0$ and again apply {\tt PSLQ}. This time, increasing precision can be done much easier. We successfully checked the result obtained in the first way with 500 digits.

As a result we obtain the following value of the missing constant $I_1(0)$ for the three-loop quark static potential
\begin{multline}
I_1(0) =
 -64 \pi ^2 \text{Li}_4\left(\frac{1}{2}\right)+6 \pi ^2 \zeta (3)+10 \zeta
   (5)-56 \pi ^2 \zeta (3) \log (2)
\\  \hspace*{-90mm}
   +\frac{53 \pi ^6}{90}-\frac{8}{3} \pi ^2
   \log ^4(2)+\frac{8}{3} \pi ^4 \log ^2(2)\;.
\label{missres}
\end{multline}
which corresponds to  $I_{18}$ in the notation of Ref.~\cite{tlpana}.

\section{Conclusions}

Using the method of differential equations we analytically evaluated the three-loop Feynman integral which was the last missing ingredient for the analytical evaluation of the three-loop quark static potential.
In the accompanying paper \cite{tlpana} this result is used to present completely analytical results for the three-loop quark static potential.
As a by-product, we obtained analytical results for some Feynman integrals typical for
the Heavy Quark Effective theory.
Our results for the master integrals involved are available\footnote{We also provide results for the $\ep$-expansion of the integrals $J_{1-9}$
defined by Eq. \eqref{hhh_MI}. Analytical results for these integrals in a format suitable for using within the package \texttt{SummerTime} can be taken
from \url{http://www.inp.nsk.su/~lee/programs/SummerTime/\#results}} in a computer-readable format at \url{http://theory.sinp.msu.ru/~smirnov/tlp}.

\paragraph{Acknowledgements} We are grateful to Matthias Steinhauser for comments on
the draft of the paper.
The work of R.L. was supported through RFBR grant No. 15-02-07893.

\bibliographystyle{JHEP}


\end{document}